# The UK and Ireland Geophysical Array – Concept and Design

Andrew Curtis (University of Edinburgh - Andrew.Curtis@ed.ac.uk – corresponding author)
Karen Lythgoe (University of Edinburgh - Karen.Lythgoe@ed.ac.uk)
Stephen P. Hicks (University College London - stephen.hicks@ucl.ac.uk)
Lidong Bie (University of East Anglia - l.bie@uea.ac.uk)
Dominik Strutz (University of Edinburgh - dominik.strutz@ed.ac.uk)
Emma Chambers (Dublin Institute for Advanced Studies - echambers@cp.dias.ie)
Brian Baptie (British Geological Survey - bbap@bgs.ac.uk)
Dave Cornwell (University of Aberdeen - d.cornwell@abdn.ac.uk)
Juliane Huebert (British Geological Survey - jubert@bgs.ac.uk)
Jessica Irving (University of Bristol - jessica.irving@bristol.ac.uk)
Glenn Jones (Cardiff University - JonesG121@cardiff.ac.uk)
Sergei Lebedev (University of Cambridge - sl2072@cam.ac.uk)
Walid Ben Mansour (Washington University in St. Louis, USA - walid@wustl.edu)
Aideliz Montiel Álvarez (University of Edinburgh, and Geological Survey of Norway - a.montiel@ed.ac.uk)
Stuart Nippress (AWE Blacknest - stuart@blacknest.gov.uk)
Koen Van Noten (Royal Observatory of Belgium - koen.vannoten@seismology.be)
Tim Pharaoh (British Geological Survey - tcp@bgs.ac.uk)
Romesh Palamakumbura (British Geological Survey - romesh@bgs.ac.uk)
Nick Rawlinson (University of Cambridge - nr441@cam.ac.uk)
Pablo Rodriguez Salgado (University College Dublin - pablo.rodriguezsalgado@ucd.ie)
James Verdon (University of Bristol - James.Verdon@bristol.ac.uk)
Chuanbin Zhu (Northumbria University - chuanbin.zhu@northumbria.ac.uk)
Wen Zhou (TNO, Netherland - wen.zhou@tno.nl)
Jelle Assink (Royal Netherlands Meteorological Institute (KNMI), The Netherlands - jelle.assink@knmi.nl)
Ian Bastow (Imperial College London - ibastow@ic.ac.uk)
Dino Bindi (GFZ Helmholtz Centre for Geosciences, Germany - dino.bindi@gfz.de)
Tom Blenkinsop (Cardiff University - BlenkinsopT@cardiff.ac.uk)
Raffaele Bonadio (University of Cambridge - rb2075@cam.ac.uk)
Michael Braim (Sproule ERCE - michael.braim@sproule-erce.com)
Tim Craig (University of Leeds - t.j.craig@leeds.ac.uk)
Elizabeth Day (Imperial College London - e.day@imperial.ac.uk)
Giovanni Diaferia (INGV, Italy - giovanni.diaferia@ingv.it)
Stuart Dunning (Newcastle University - stuart.dunning@ncl.ac.uk)
Ben Edwards (University of Liverpool - ben.edwards@liverpool.ac.uk)
Ake Fagereng (Cardiff University - FagerengA@cardiff.ac.uk)
Stewart Fishwick (University of Leicester - sf130@leicester.ac.uk)
Amy Gilligan (University of Aberdeen - amy.gilligan@abdn.ac.uk)
David Green (AWE Blacknest - dgreen@blacknest.gov.uk)
David Healy (University of Leeds - D.Healy@leeds.ac.uk)
Anna Horleston (University of Bristol - Anna.Horleston@bristol.ac.uk)
Mark Ireland (Newcastle University - Mark.Ireland@newcastle.ac.uk)
Jenny Jenkins (Durham University - jennifer.jenkins@durham.ac.uk)
Jessica Johnson (University of East Anglia - Jessica.Johnson@uea.ac.uk)
Mike Kendall (University of Oxford - mike.kendall@earth.ox.ac.uk)

Tom Kettlety (University of Oxford - tom.kettlety@earth.ox.ac.uk)
Duygu Kiyan (Dublin Institute for Advanced Studies, Dublin, Ireland - duygu@cp.dias.ie)
Paula Koelemeijer (University of Oxford - paula.koelemeijer@earth.ox.ac.uk)
Rita Kounoudis (University of Oxford - rita.kounoudis@earth.ox.ac.uk)
Victoria Lane (University of Leicester - vl36@leicester.ac.uk)
Chuanchuan Lu (University of Cambridge - chuanchuanlu@esc.cam.ac.uk)
Alan MacDonald (British Geological Survey – amm@bgs.ac.uk)
Fabrizio Magrini (ANU, Australia - fabrizio.magrini@anu.edu.au)
Auggie Marignier (University of Oxford - auggie.marignier@earth.ox.ac.uk)
Carl Martin (Australian National University - Carl.Martin@anu.edu.au)
Martin Möllhoff (Dublin Institute for Advanced Studies – martin@dias.ie)
Iain Neill (University of Glasgow - iain.neill@glasgow.ac.uk)
Andy Nowacki (University of Leeds - a.nowacki@leeds.ac.uk)
Bob Paap (TNO, Netherlands – bob.paap@tno.nl)
Simone Pilia (KFUPM, Saudi Arabia - simone.pilia@kfupm.edu.sa)
Sjoerd de Ridder (University of Leeds - S.Deridder@leeds.ac.uk)
Elmer Ruigrok (Royal Netherlands Meteorological Institute (KNMI), The Netherlands - elmer.ruigrok@knmi.nl)
Peidong Shi (University of Cardiff - ShiP1@cardiff.ac.uk)
Anna Stork (AtkinsRéalis - anna.stork@atkinsrealis.com)
Alice Turner (University of Cambridge - art77@cam.ac.uk)
Jim Whiteley (AtkinsRéalis - Jim.Whiteley@AtkinsRealis.com)
Anton Ziolkowski (University of Edinburgh - Anton.Ziolkowski@ed.ac.uk)

*Scientific exploration of the UK and Ireland's subsurface has made important contributions to scholarship and prosperity for people and the planet, including economic growth, sustainable use of natural resources, storage of greenhouse gases, and inspiring curiosity about the Earth beneath our feet. This article outlines a vision for an array of seismological instruments spanning the UK and Ireland (UKI-Array), augmented by other types of geophysical sensors, to maximise the value offered by existing equipment pools. The mission is to research natural phenomena and structure in the deep and shallow Earth, to solve problems concerning hazards and resources, to connect scientists to schools and the broader public, and thus to inspire a new generation to learn about geophysics. The vision was created through a community driven process of engagement and participation. This paper describes the concept and design of the UKI-Array; a companion paper discusses related opportunities and potential applications.*

# Introduction

The UK and Ireland are currently situated within a tectonically stable continental region, with no active volcanoes nor major fault ruptures. Yet, their shared subsurface has consistently proven to be both fascinating and critically important. Subsurface hazards such as moderately sized earthquakes pose a substantial risk to critical engineering sites (e.g., Tromans et al., 2019; Mosca et al., 2022; Lebedev et al. 2023). In turn, subsurface engineering projects may trigger earthquakes (Verdon et al., 2025) or risk contaminating water resources (Stuart 2011; Smedley et al., 2024). Subsurface material properties can vary under the influence of stress, temperature or precipitation changes which lead to landslides or mass movements (Pennington et al., 2015). The subsurface is also a crucial source of natural resources such as water, rock, soil, heat, minerals and hydrocarbons (Allen et al., 1997; Deady et al., 2023; Murtaza et al. 2021; Patton et al. 2025; Scanlon et al., 2023). It is a natural storage option for compressed-air, gas or hydrogen-based energy storage (Evans et al., 2018; Williams et al., 2022), of excess heat for later release (Reuss 2015; Fraser-Harris et al., 2022), and for waste products such as spent nuclear fuel (Tweed et al., 2015; Pavey et al., 2025) and the vast quantities of $CO_2$ that must be locked away from the atmosphere to mitigate future climate change (Haszeldine, 2012). And imaging the Earth's subsurface provides critical clues as to its geological history and the tectonic processes that shaped it (Bonadio et al., 2021; Galetti et al., 2017; Nicolson et al., 2014).

Each of these perspectives on the subsurface caries social, engineering or economic risks related to the geology and variability of Earth dynamics, yet each risk of natural or engineered hazards comes with opportunities for mitigation through the use of scientific knowledge and observational data. Obtaining information about the subsurface can involve operations such as drilling or excavation, which are expensive and invasive and only cover a limited volume. It is therefore important to be able to explore, image and monitor changes in the subsurface, at a wide range of scales, remotely.

The primary source of remotely sensed information is geophysical. Geophysicists use measurements of the seismic, electrical, electromagnetic and gravitational energy (amongst others), to infer properties of the Earth's surface or interior. In particular, seismic waves from active sources, from earthquakes, and from more distributed energy sources (referred to as ambient seismic noise), contain information about both the sources themselves and the structure and properties of the Earth. As a set of islands, the UK is bombarded from all directions by ambient noise created by oceanic waves. By analysing recordings of these waves and other signals recorded on dense arrays of sensors, geophysicists can construct 3-D images of crucial properties of the subsurface. Monitoring changes over time reveals processes active inside the Earth across multiple scales of time and space.

Thanks to the seismic stations operated by the British Geological Survey (BGS) and AWE Blacknest in the UK, and the Irish National Seismic Network (INSN) in Ireland (BGS 1970; Blacknest 1960; DIAS 1993), and to the recent temporary broadband station deployments in Ireland, the lithosphere and underlying mantle beneath large parts of the United Kingdom, Ireland, and surroundings can now be imaged (e.g., Bonadio et al. 2021), which can help us to de-risk geothermal energy exploration (Chambers et al. 2025), better assess subsurface mineral resources (e.g., Roy et al. 2026), and determine what controls the distributions of seismicity and seismic hazard (Mosca et al., 2022). However, substantial gaps in the station coverage across Britain and Ireland remain, and until now, attempts to image the crust and mantle beneath the region have been unable to resolve details less than several tens of kilometres across (Nicolson et al., 2014; Luckett & Baptie, 2015; Bonadio et al., 2021; Zhao et al., 2026). This poor resolution is due to the limited number of sensors in the permanent seismic network, and this sparsity similarly limits our ability to detect and locate small natural earthquakes and other seismic sources, for example to attribute them to specific fault zones and use them to alert us to the possibility of larger future events (Roy et al., 2021; Watkins et al., 2023), or to distinguish between natural and anthropogenic causes (e.g., Hicks et al., 2019).

Currently, the UK has a large number of seismic sensors (see Table 1), many of which were acquired recently to form the LeNS-UK community facility, funded by the Natural Environment Research Council (NERC). While subsets of these have been used successfully for smaller, bespoke projects related to the goals of individual research groups (e.g., Hudson et al., 2024), it is also timely to consider whether an integrated ambition to combine these sensors with other existing pools of equipment in the UK and Ireland to tackle combined objectives could lead to scientific and technical advances that would benefit the UK Earth science and other related communities.

Complementary to seismic imaging, magnetotellurics (MT) is a deep sounding passive electromagnetic method that images electrical properties (influenced by, e.g., the presence of fluids, melt, metallic minerals) to great depth (1- 1000 km). Backbone, long-period MT data can be used to investigate lithospheric structure, space weather impacts, and characteristics of specific targets such as fault zones, and for geothermal and mineral exploration (e.g., lithium in geothermal brines hosted in fractured rocks and sulphide deposits – see Hübert et al., 2025). And other electromagnetic and electrical methods allow the top kilometre or so of the Earth's subsurface to be interrogated in greater detail.

Adding acoustic sensors to seismometer locations allows airborne infrasound waves to be recorded. Infrasound recorded across a network of such sensors can assist in improving our understanding of dynamic atmospheric process including middle-atmosphere wind variability. When combined with seismic recordings, infrasound data can also provide valuable information about the partitioning of energy between the subsurface and atmosphere for near-surface events (Greene et al., 2009).

The concept of such a combined collection of sensors is referred to here as the *UK and Ireland Array* (*UKI-Array* for short). In order to test and develop a community vision for the UKI-Array, a small team (the first five authors) created an initial draft of a green paper (a conceptual proposal), and through the British Geophysical Association's email list (https://geophysics.org.uk/sign-up/), invited the whole UK Geophysical community to comment and add potential opportunities, applications and implications of such an array. The team consolidated these contributions in the green paper, then organised a hybrid Discussion Meeting in London and online, hosted by the Royal Astronomical Society on October 10th, 2025. They issued an open invitation to the geophysical community to use the meeting to discuss the vision, the array's design, how the data might inform the wide range of interests of the community, and to voice concerns about deployment, funding and a range of other aspects. There was a strongly positive response to the UKI-Array concept, a wide range of proposed applications and outreach opportunities from the array deployment itself, and a clear signal from the community to move this concept forward based on proposals in the green paper and subsequent discussion. These proposals and discussions are summarised here to form a consolidated description of the UKI-Array initiative and its potential applications.

In this article, we first describe the UKI-Array concept in more detail. Then a snapshot of the equipment available across the UK and Ireland is listed, and conceptual end-member designs are discussed. In the companion paper Curtis et al., (2026) we describe the various ideas and opportunities, namely: to explore surface processes and environmental geophysics; to image the subsurface of the UK and Ireland landmass and surrounding seas; to monitor and model earthquakes, hazard and risk; to investigate potential Earth resources and subsurface storage; and finally to illuminate structure and properties of the deeper Earth.

## UKI-Array concept

The UKI-Array concept comprises the deployment of thousands of seismic sensors across the UK and Ireland. This baseline seismological array may be augmented by seismometers deployed on the seabed and in neighbouring countries across the English Channel and North Sea, and by a range of other complementary geophysical sensor types. The resulting data will provide detailed models of the Earth's crust under the UK and Ireland, high-resolution images of key fault zones, and, over the deployment period, information about both natural and anthropogenic seismic activity, including low-magnitude earthquakes commonly associated with industrial or other anthropogenic activity. The array will be designed to answer fundamental scientific questions about the shallow and deep Earth, and to address important issues relating to the future use of the Earth's subsurface both as a source for sustainable energy and resources, and as a means of energy and waste storage. In addition to subsurface applications, parts of the network could be used for near-surface environmental applications, such as the close-range sensing of properties that control geomorphic processes – landslides, floods, water table changes, and sediment transport (Bainbridge et al. 2022). The data would be openly available, and a significant component of the activity would focus on public engagement.

The existing network of permanent seismic sensors reliably detects events of magnitude greater than M ~3.0 anywhere in the UK, while in Ireland this so-called magnitude of completeness is M ~1.2 (Baptie 2018; Mosca et al., 2022; Möllhoff and Bean 2016; Möllhoff et al., 2019; Musson & Sergeant, 2007). Yet, tens of thousands of smaller magnitude earthquakes remain undetected each year. There is a pressing need to enhance our ability to detect small earthquakes to facilitate the monitoring, derisking, and regulation of industrial activities in the upcoming era of subsurface waste and energy storage, and sustainable resource production (Karamzadeh et al., 2021) . The short-term deployments in the UKI-Array would therefore be designed in part to provide significantly updated images and information about the UK's subsurface structure and faults, as well as offering an initial test of the monitoring capability of a dense, optimally-designed array of heterogeneous sensor types.

## Geophysical equipment

Table 1 lists the seismological sensors held in institutions around the UK and Ireland. Ocean bottom seismometers are also available (including ~60 at the Ocean Bottom Instrumentation Consortium (OBIC) arm of the NERC-funded Geophysical Equipment Facility (GEF). These are not listed in Table 1 because deployment of seabed sensors is relatively expensive, so they have not been added to the array designs presented below. If the opportunity to include them materialises then designs will be updated. A number of Raspberry Shake seismometers have also not been included because these are typically deployed in people's houses, so their locations are not shared publicly. Nevertheless, with appropriate data quality checks they can significantly enhance earthquake catalogues by detecting small events that would otherwise go unnoticed by national networks alone, and their low cost makes them valuable tools for public outreach.

A significant portion of the available equipment is provided by LeNS-UK – a NERC-funded instrumental facility that was established in September 2025. LeNS-UK hosts two types of seismic nodes (Figure 1).

There are 1,980 Stryde Range+ nodes, which are small, compact instruments containing a single component accelerometer that can record for 50 days, and 200 SmartSolo nodes, which contain 3-component geophones with a longer period response and can record for 30 days.

In the UK and Ireland seismological community, there are also smaller but numerous pools of broadband seismometers, which record to a much lower frequency response, and can detect long-period teleseismic surface waves and enable full-waveform source and structure imaging, for example. Broadband seismometers, solar panels, and support can also be provided by NERC's SEIS-UK facility, which is part of NERC's Geophysical Equipment Facility (GEF). SEIS-UK hosts a large number of broadband instruments for community use, which will strongly complement the denser short-period node networks.

The large volumes of seismological data (likely to total tens to hundreds of terabytes) collected would be stored on the British Geological Survey's EIDA storage and retrieval node. EIDA is the European Integrated Data Archive network, which shares seismic data between data provision and compute servers internationally. This will ensure that the data are widely available in standard formats, according to the FAIR Data Principles (Findable, Accessible, Interoperable, Reusable), and can be run through standard as well as bespoke processing sequences.

Recently, the NERC GEF acquired 10 new broadband MT systems, providing the community with this much-needed infrastructure, while the British Geological Survey (BGS) operates several additional instruments. Similarly, the Dublin Institute for Advanced Studies (DIAS) MASTER Infrastructure hosts 12 ultra-wide band MT systems and 20 long-period MT systems. Additional types of geophysical instrumentation that probes the electrical properties of the subsurface are of potential interest at smaller spatial scales. Several universities and companies in the UK own Electrical Resistivity Tomography (ERT) equipment, used to image the electrical conductivity structure of the upper tens of metres of the crust. Active-source electromagnetics has also been suggested: one such system that uses a current bipole source and electric field receivers can penetrate to depths of around two kilometres, and the NERC Geophysical Equipment Facility has a transient electromagnetic instrument (TEM), both systems being hosted at the University of Edinburgh.

Fibre-optic based, distributed acoustic sensing (DAS) systems measure dynamic strain, such as that caused by seismic waves, along the one-dimensional profiles of optical fibres laid on the surface or buried. The key advantage of these systems is that they can be used to record seismic energy at spatial intervals of the order of 1 to 10s of metres along a cable length of up to approximately 100 km. Although burying cables can be time consuming and logistically challenging and DAS data volumes tend to be huge, several systems are available in the UK and Ireland (including one at the University of East Anglia that is a NERC-funded community asset) which may be used to complement either dense local arrays or large-scale linear arrays, and it is also possible to connect DAS interrogators to existing subsurface fibre that was originally deployed for other purposes such as telecommunications.

***Table 1:*** *Seismological instrumentation held at institutions around the UK and Ireland. (BB: broadband).*

| Institution | Equipment |
| --- | --- |
| **Dublin Institute for Advanced Studies** | • 65 3C SmartSolo<br>• 134 1C SmartSolo<br>• 20 broadband<br>• 2 DAS interrogators |
| **University of Oxford** | • 21 Raspberry Shakes<br>• 280 STRYDE nodes<br>• 6 broadband<br>• 13 Güralp Certimus BB seismometers + solar panels<br>• 420 Sercel WiNG DFUs (1C BB MEMS)<br>• 10 mounting plates for 3C configuration (Galperin)<br>• 1 DAS interrogator (Sintela) |
| **University of Aberdeen** | • 6 Güralp Certimus<br>• 10 Güralp 6TD<br>• 102 3C SmartSolo nodes<br>• 20 Raspberry Shakes |
| **Birkbeck** | • 3 Raspberry Shakes |
| **University of East Anglia** | • 1 Güralp 40T<br>• 2 Raspberry Shakes<br>• 1 DAS interrogator |
| **Strathclyde University** | • 7 3C SmartSolo |
| **University of Leeds** | • ~50 3C SmartSolo<br>• 1 DAS interrogator (Febus) |
| **University of Edinburgh** | • 21 3C SmartSolo |
| **University College London** | • 5 Güralp broadband quick deploy kits |
| **British Geological Survey** | • ~20 broadband<br>• EIDA system to store data |
| **University of Cambridge** | • 5 x 120s Trilliums |
| **University of Bristol** | • 4 Güralp Certimus<br>• 6 Trilliums<br>• 1 DAS interrogator |
| **Cardiff University** | • 3 Güralp Radian |
| **British Antarctic Survey** | • 32 3C SmartSolo<br>• 2 Güralp Radians |
| **Seis-UK (University of Leicester)** | • Many broadbands + dataloggers + solar panels |
| **LenS-UK (University of Cambridge)** | • 200 3C SmartSolo<br>• 1980 1C Stryde nodes (~50 days recording time) |
| **Imperial College London** | • 1 Güralp ESP |
| **SEIS-UK Equipment Pool** | • 189 broadband (various makes/models)<br>• 37 High Frequency Seismometers (36 Polar Pegasus; 1 Geode)<br>• 20 Pegasus & Trillium Compact (NERC Urgency Pool) |

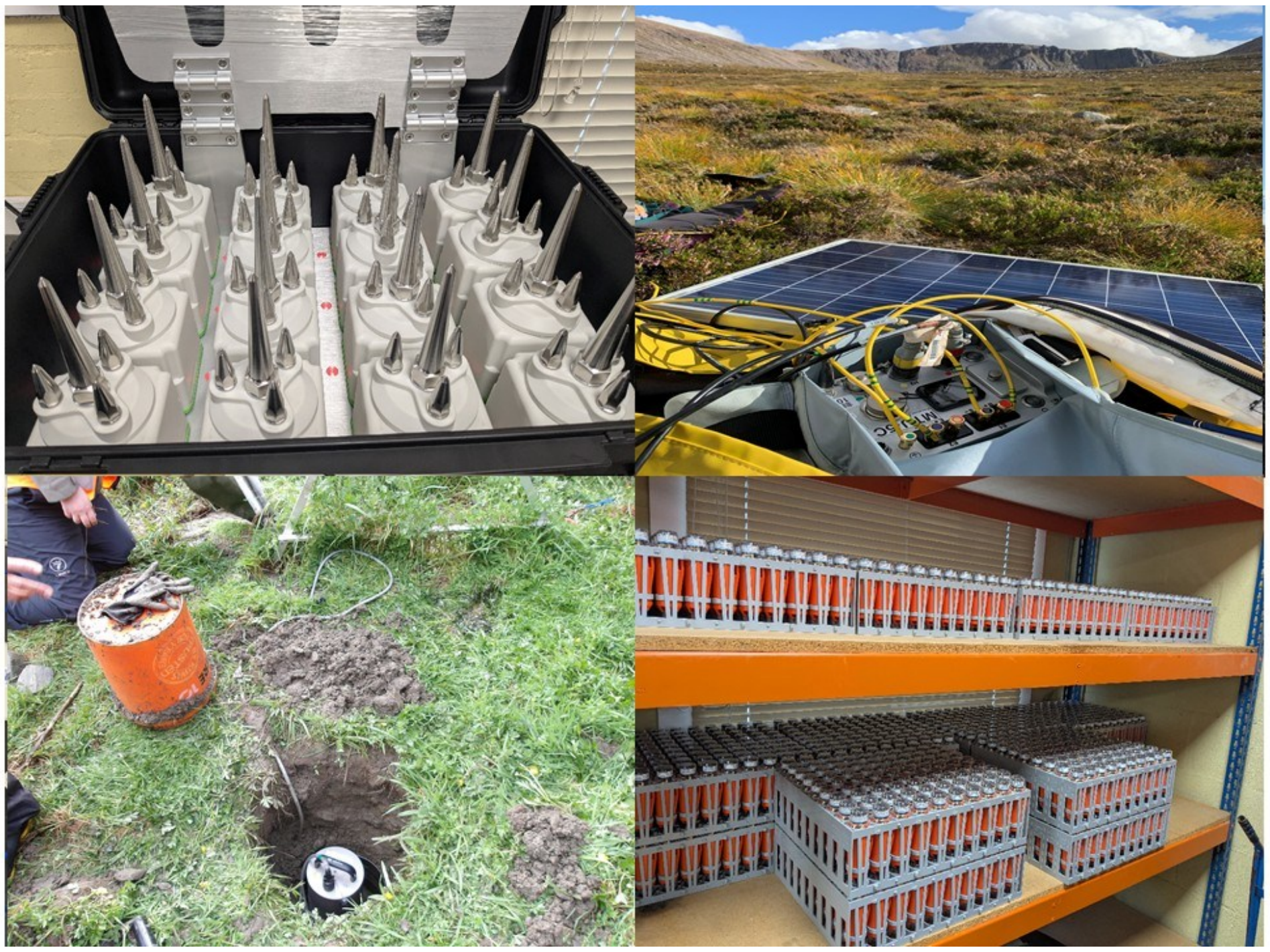

***Figure 1**: Top left: SmartSolo node battery packs under charge; Bottom left: A broadband seismometer being deployed, ready to be covered then buried; Bottom right: A subset of the 1,980 Stryde nodes at the LeNS-UK facility; Top right: Phoenix MTU-5C ultra-wideband magnetotelluric receiver close to Coire an t-Sneachda. Photo credits: co-authors Nick Rawlinson (top-left, bottom-right) and Juliane Hübert (top-right), and Jack Andrew Smith, University of Edinburgh (bottom-left).*

## Design Components and potential targets

The UKI-Array array concept consists of one or more of the following six components:

1. **A distributed network of >2,000 seismic sensors across the UK and Ireland** (a concept design is shown on the right of Figure 2), with further sensors possibly deployed offshore.

This network would remain in place for at least two months. Several studies suggest that one month (the battery life of SmartSolo nodes) is sufficient time to obtain both stable cross-correlations for imaging with ambient noise and to estimate coherencies in spatial and temporal local wavefield gradients (Nicolson et al., 2014; Cao et al., 2020). This would need to be checked for higher frequency Stryde nodes, and the feasibility of longer deployment times for nodes can be considered. This component would deliver a snapshot of seismicity across the UK and Ireland over the survey period, providing the first uniformly dense catalogue of earthquake locations (Figure 2(d)). It would produce new high-resolution, homogeneous images of the crust and upper mantle from surface and body waves. Shallow crustal models would have commercial value, while improved knowledge of faults and stresses would help to identify areas at higher risk of induced seismicity from future subsurface activities. The dataset would support studies of deeper Earth processes of broad national and international interest, and constrain crustal seismic attenuation, improving magnitude estimates for small, induced earthquakes and thus aid regulation of industrial activity.

2. **Clustered, potentially transportable arrays involving subsets of sensors** deployed more densely around specific targets of interest, such as to test scientific hypotheses, or to image or monitor potentially active faults, landslide/flood risk hotspots, and geological targets with economic value.

The concept is illustrated in Figure 3(c), focussing on granites of geothermal interest in the south-west UK, potentially coal mine hazards such as in South Wales that also create opportunities for subsurface heat storage and extraction, and areas of potential landslides or mass movements in Scotland and northern England. This component also includes small linear arrays to characterise major fault lines that segment the UK and Ireland along north-easterly trending boundaries. It would generate high-resolution images of selected shallow and deep crustal and upper-mantle targets across the UK and Ireland using surface and body waves. It would allow for detection and characterisation of hazardous surface processes in noisy environments. A focused sub-array in southeast England, for example, would address the London monitoring gap and establish baseline data for subsurface resource exploration and future production, and for subsurface storage.

3. **A linear array** stretching approximately north-south across the length of the UK, that crosses major ancient tectonic suture zones (Figure 3(c)).

This array would broadly follow a 1974 active-source seismic survey (Bamford et al., 1976) with the aim to clarify relationships between major geological boundaries. The UK and Ireland are dominated by NE–SW and E–W orogenic belts formed during successive mountain-building events, but key aspects of their tectonic evolution, such as past subduction zone locations, remain debated. An extremely dense linear array would provide an unbiased transect of information interrogating the properties of known geological boundaries, faults with less obvious surface expression, and the heterogeneous material characteristics of intervening regions.

4. **An additional network of temporary broadband seismic stations** (or re-deployment of existing stations in new locations) deployed for two or more years to ensure successful application of a range of passive seismic imaging techniques (illustrated in Figure 2(a) and (b) for a conservative addition of only 20 broadband sensors). The new locations will fill gaps in the broadband-station coverage provided by the current configurations of the permanent seismic networks in the UK and Ireland.

This broadband network would allow the seismic and thermal structure of the crust and underlying mantle to be imaged, with implications for the distributions of seismicity, seismic hazard, geothermal energy and mineral resources. Broadband seismometers play a major role in larger magnitude earthquake characterisation, and this capability would improve across most of the region.

5. **Other types of sensors** may be added to any of these campaigns, for example, to probe gravitational, electrical, magnetic, electromagnetic, or acoustic effects and properties of the Earth's (sub)surface and atmospheric processes.

This component extends the seismological backbone array components to the wider geophysical community, enabling shared deployment logistics (e.g., co-location may allow power supplies and land access rights to be shared). It would support MT studies at scales similar to Component 1, complements Component 2 with electrical resistivity and active-source electromagnetics, and may include acoustic sensors and fibre-optic distributed acoustic sensing alongside selected seismic arrays.

6. **Coordinated synchronous arrays on opposite coasts of the English Channel, North Sea, and Irish Sea** would allow offshore subsurface structures to be imaged (Figure 4) and offshore seismicity to be better characterised. Ocean bottom seismometers from OBIC (UK) and DIAS iMARL (Ireland) pools could be used to significantly improve off-shore seismicity monitoring and subsurface imaging.

Component 6 extends the array offshore to support seismic monitoring of $CO_2$ sequestration in the Southern North Sea (e.g., TNO 2025), which has a fairly high seismicity rate compared to on-land Britain. Accurate event locations require high-resolution velocity models (Verdon et al., 2013; Strutz & Curtis 2025), but S-wave structure remains poorly constrained (Jerkins et al., 2025). Coordinated coastal deployments across northwest Europe and the UK would enable ambient noise tomography to refine deep velocity models and strengthen offshore monitoring.

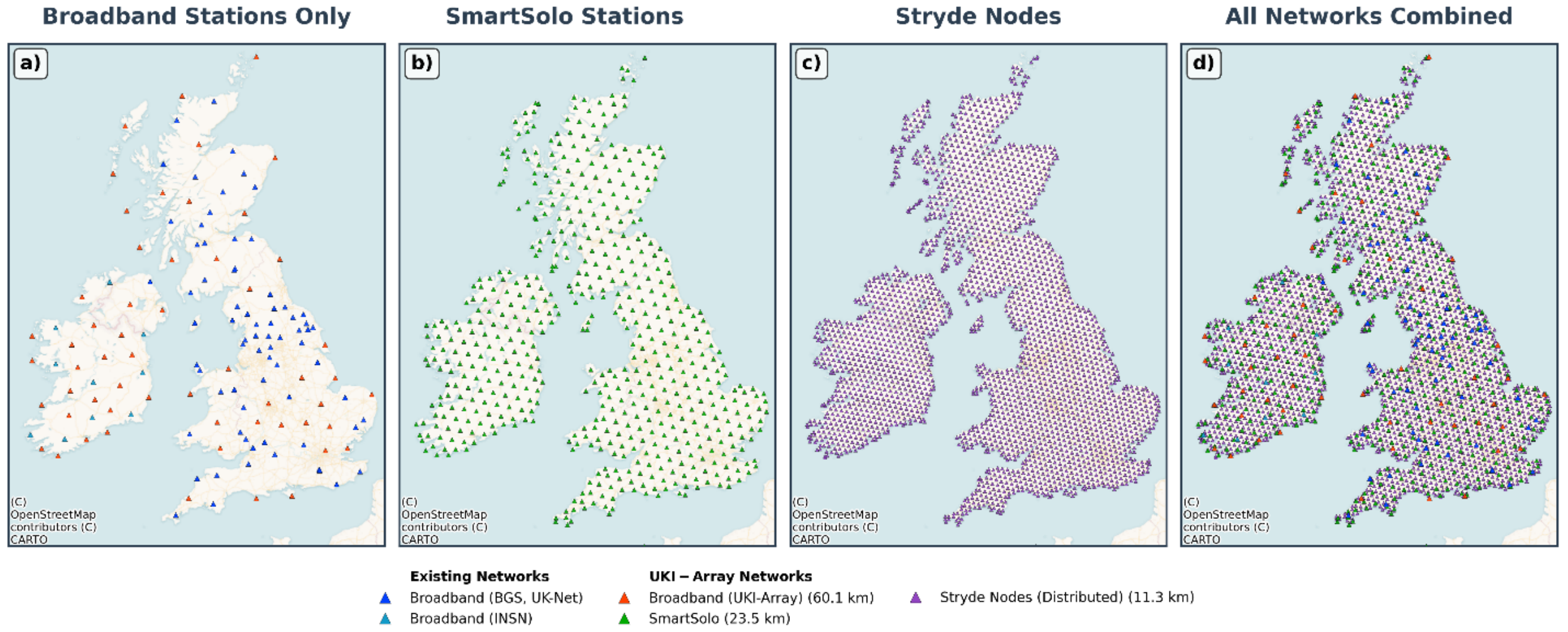

***Figure 2:*** *Illustration of sensor density and spacing achieved if existing broadband stations are combined with SmartSolo and Stryde seismic nodes distributed evenly across the UK and Ireland. The average inter-station distances are shown in brackets in the legend. SEIS-UK equipment has not been included.*

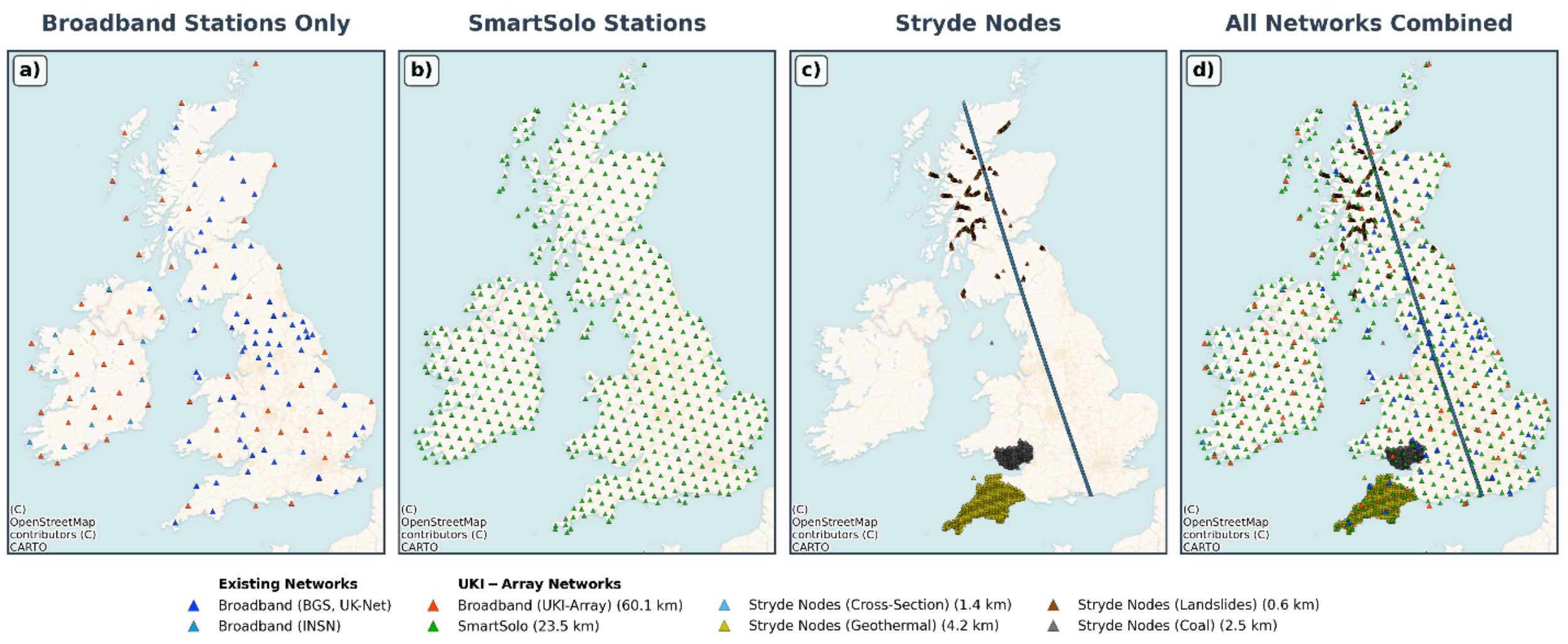

***Figure 3:*** *Illustration of an array in which the current broadband equipment and SmartSolo nodes in Figure 2 are still distributed according to design component 1 (evenly spread), while all Stryde nodes are used for design component 2 (focussed sub-arrays around targets of interest). The average inter-station distances are shown in brackets in the legend.*

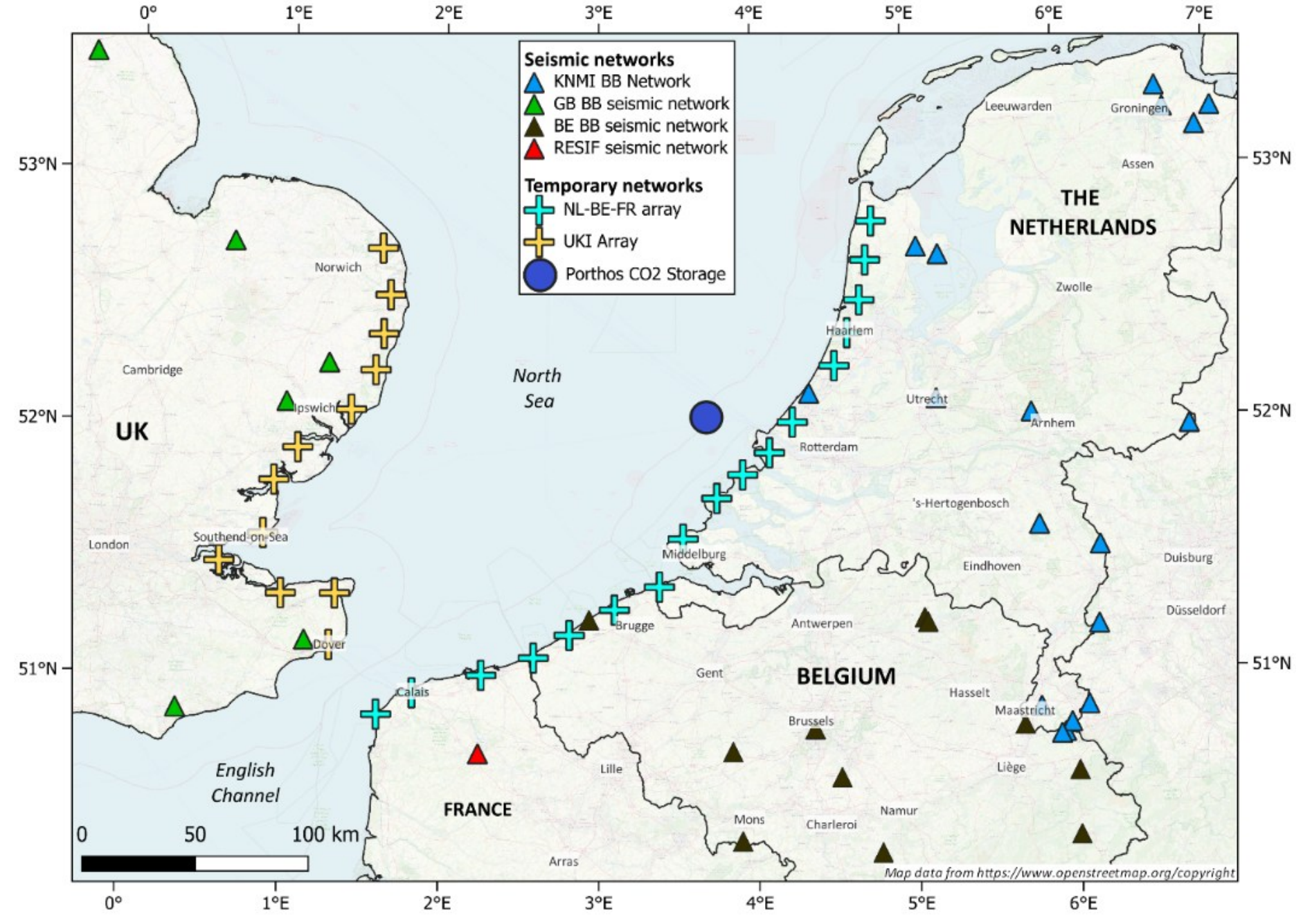

***Figure 4:*** *Conceptual design of an international component of the UKI-Array concept. In collaboration with Dutch, Belgian and French seismologists, the array including low-frequency nodes and seismometers could improve North Sea monitoring to image structures of scientific and economic interest beneath the offshore seabed surrounding the UK and Ireland, such as around the Porthos CO2 storage site (circled) or around new windfarms or energy islands in the North Sea.*

## Discussion

In this first of two companion papers we have set out the overall concept of the UKI-Array, and various design components. A variety of other considerations must be integrated to create a final, specific design. Our conceptual designs were primarily developed to estimate average inter-station distances given the equipment that is readily available. Ultimately these designs must be optimised jointly to maximise both coverage and resolution of tomographic imaging, and earthquake monitoring efficacy, given potential land access constraints (Curtis et al., 2004; Mordret & Grushin, 2025; Strutz et al., 2025).

It would be useful to test the performance of Stryde nodes at long inter-node distances and low-frequencies, both as single-component (vertical) sensors, or when operating as 3-component seismometers to record shear waves (using three orthogonally oriented nodes per location). For ambient noise seismic interferometry and imaging, the same noise field must be recorded by pairs of sensors at different locations. Ambient noise tomography is therefore affected by the degree of seismic wave attenuation, and by the spatial coherency of ambient seismic noise across the UK and Ireland. While a previous test has already shown that ambient noise imaging works well with Stryde nodes in either 1- or 3-component mode, up to at least 6 km inter-node spacing in Cornwall, this was only possible because the manufacturer removed the nodes' standard low-frequency cut-off. Local test deployments in other regions would therefore be a good way to quantify uncertainties before designing the final array.

The array(s) will require large-scale deployment, servicing, and recharging across much of the UK and Ireland, with different sensor types operating at national and sub-array scales. Beyond the clear educational benefits, engagement with the public and schools will also help with site hosting, data use and education, requiring substantial organisation. Fair fieldwork contributions from all partners across different regions of the UK and Ireland, and joint effort to obtain site permitting, are also important. One idea would be to promote the project prior to deployment so that members of the public can register their interest in hosting a sensor, helping to ensure secure deployment.

A challenge that may constrain deployment of the various components of the array is that relatively few sources of science council funding allow fair distribution of resources between the UK and Ireland. This is a significant barrier to deploying multi-national arrays more generally, and may require that the UK and Irish parts of the UKI-Array are deployed during offset time periods, or with different levels and types of resourcing. This will require that the survey and logistical design methods are flexible and responsive to changing opportunities, as they arise.

In this paper we have discussed the concept and design of the UKI-Array, and some of its benefits. In the companion paper Curtis et al., (2026), we explore the specific applications and opportunities enabled by the dataset in more detail. This reveals the value offered by such an initiative, contributing to a justification of the effort and expense required. Together these papers form a coherent argument for the community to progress this initiative to a funding proposal, and to eventual implementation.

## Acknowledgements

The authors would like to thank the British Geophysical Association for help initiating this community activity, and the Royal Astronomical Society for sponsoring the Discussion Meeting held at the society's premises on 10th October 2025 which contributed greatly to the coordination of this article.

## References

Allen, D.J., Brewerton, L.J., Coleby, L.M., Gibbs, B.R., Lewis, M.A., MacDonald, A.M., Wagstaff, S.J., and Williams, A.T., 1997. The physical properties of major aquifers in England and Wales. British Geological Survey Technical Report WD/97/34. 312pp. Environment Agency R&D Publication 8.

Bainbridge, R., Lim, M., Dunning, S., Winter, M. G., Diaz-Moreno, A., Martin, J., … Jin, N. (2022). Detection and forecasting of shallow landslides: lessons from a natural laboratory. Geomatics, Natural Hazards and Risk, 13(1), 686–704.

Bamford, D., Faber, S., Jacob, B., Kaminski, W., Nunn, K., Prodehl, C., ... & Willmore, P. (1976). A lithospheric seismic profile in Britain—I preliminary results. Geophysical Journal International, 44(1), 145-160.

BGS 1970. British Geological Survey. Great Britain Seismograph Network [Data set]. International Federation of Digital Seismograph Networks. https://doi.org/10.7914/av8j-nc83

Baptie, B., 2018. Earthquake Seismology 2017/2018. British Geological Survey Open Report, OR/18/029

Blacknest, 1960. UK-Net, Blacknest Array [Data set]. International Federation of Digital Seismograph Networks. https://doi.org/10.7914/nz1t-5w85

Bonadio, R., Lebedev, S., Meier, T., Arroucau, P., Schaeffer, A.J., Licciardi, A., Agius, M.R., Horan, C., Collins, L., O'Reilly, B.M. and Readman, P.W., 2021. Optimal resolution tomography with error tracking and the structure of the crust and upper mantle beneath Ireland and Britain. Geophysical Journal International, 226(3), 2158-2188.

Cao, R., Earp, S., de Ridder, S., Curtis, A., Galetti, E., 2020. Near-real time near-surface 3D seismic velocity and uncertainty models by wavefield gradiometry and neural network inversion of ambient seismic noise. Geophysics, 85(1), pp.KS13-KS27

Chambers, E.L., Fullea, J., Kiyan, D., Lebedev, S., Bean, C.J., Meere, P.A., Daly, J.S., Willmot Noller, N., Raine, R., Blake, S. and O'Reilly, B.M., 2025. A new subsurface temperature model for Ireland from joint geophysical–petrological inversion of seismic, surface heat flow and petrophysical data. Geophysical Journal International, 243(1), p.ggaf281.

Curtis, A., Michelini, A., Leslie, D., Lomax, A., 2004. A deterministic algorithm for experimental design applied to tomographic and microseismic monitoring surveys. Geophysical Journal International, 157(2), 595-606

Curtis, A., Lythgoe, K., Hicks, S.P., Bie, L., Strutz, D., Chambers, E., et al., 2026. The UK and Ireland Geophysical Array – Opportunities and Applications. Upcoming, in journal Astronomy and Geophysics.

Deady, E., Goodenough, K.M., Currie, D., Lacinska, A., Grant, H., Patton, M., Cooper, M., Josso, P., Shaw, R. A., Everett P., Bide, T., 2023. Potential for Critical Raw Material Prospectivity in the UK. British Geological Survey report CR/23/024.

DIAS 1993. Dublin Institute for Advanced Studies. Irish National Seismic Network [Data set]. International Federation of Digital Seismograph Networks. https://doi.org/10.7914/SN/EI

Evans, D. J., Carpenter, G., Farr, G., 2018. Mechanical Systems for Energy Storage – Scale and Environmental Issues. Pumped Hydroelectric and Compressed Air Energy Storage. In, Energy Storage Options and Their Environmental Impact, ed. R. E. Hester and R. M. Harrison, The Royal Society of Chemistry, pp. 42-114.

Fraser-Harris, A., McDermott, C. I., Receveur, M., Mouli-Castillo, J., Todd, F., Cartwright-Taylor, A., Gunning, A., and Parsons, M., 2022. The geobattery concept: a geothermal circular heat network for the sustainable development of near surface low enthalpy geothermal energy to decarbonise heating. Earth Science, Systems and Society, 2 (1), 10047.

Galetti, E., Curtis, A., Baptie, B., Jenkins, D., and Nicolson H., 2017. Transdimensional Love-wave tomography of the British Isles and shear-velocity structure of the East Irish Sea Basin from ambient-noise interferometry. Geophysical Journal International, 208(1), 36–58

Green, D.N., Guilbert, J., Le Pichon, A., Sebe, O., and Bowers, D., 2009. Modelling Ground-to-Air Coupling for the Shallow $M_L$ 4.3 Folkestone, United Kingdom, Earthquake of 28 April 2007. Bulletin of the Seismological Society of America, 99 (4), 2541–2551.

Haszeldine, S., 2012. UK carbon capture and storage, where is it? Energy & Environment, 23 (2/3), Special Issue: Carbon Dioxide Capture and Storage (CCS), pp. 437-450.

Hübert, J., Eaton, E., Beggan, C.D., Montiel-Álvarez, A.M., Kiyan, D., and Hogg, C., 2025. Developing a new ground electric field model for geomagnetically induced currents in Britain based on long-period magnetotelluric data. Space Weather, 23, e2025SW004427.

Hudson, T.S., Kettlety, T., Kendall, J.M., O'Toole, T., Jupe, A., Shail, R.K. and Grand, A., 2024. Seismic node arrays for enhanced understanding and monitoring of geothermal systems. The Seismic Record, 4(3), pp.161-171.

Jerkins, A. E., Schweitzer, J., Kettlety, T., Evgeniia Martuganova, Kühn, D., and Oye, V., 2025. Relocating seismic events in the North Sea: challenges and insights for earthquake analysis. *Geophysical Journal International*, *241*(1), 728–742.

Karamzadeh, N., Lindner, M., Edwards, B.. Gaucher, E., and Rietbrock, A., 2021. Induced seismicity due to hydraulic fracturing near Blackpool, UK: source modeling and event detection. J Seismol 25, 1385–1406.

Lebedev, S., Grannell, J., Arroucau, P., Bonadio, R., Agostinetti, N.P. and Bean, C.J., 2023. Seismicity of Ireland, and why it is so low: How the thickness of the lithosphere controls intraplate seismicity. Geophysical Journal International, 235(1), 431-447.

Luckett, R., Baptie, B., 2015. Local earthquake tomography of Scotland, Geophysical Journal International, 200(3), 1538–1554.

Möllhoff, M., & Bean, C.J., 2016. Seismic Noise Characterization in Proximity to Strong Microseism Sources in the Northeast Atlantic, Bulletin of the Seismological Society of America, 106(2), 464–477.

Möllhoff, M., Bean, C.J. & Baptie, 2019. B.J. SN-CAST: seismic network capability assessment software tool for regional networks-examples from Ireland. Journal of Seismology, 23, 493–504.

Mordret, A., Grushin, A. G., 2025. Beating the aliasing limit with aperiodic monotile arrays. Phys. Rev. Appl., 23 (3), 034021.

Mosca, I., Sargeant, S., Baptie, B., Musson, R. M. W., & Pharaoh, T. C., 2022. The 2020 national seismic hazard model for the United Kingdom. Bulletin of Earthquake Engineering, 20, 633–675.

Murtaza, A., Mahnoor, M., Inam, A., Shah, A.A., Obeidi M.A., & Ahad, I.U.I., 2021. Critical Raw Materials for Ireland for a Resource-Efficient Circular Economy (CIRCLE), EPA Research Report 2021-GCE-1046, Report No. 478

Musson, R. M. W., Sargeant, S. L., 2007. Eurocode 8 seismic hazard zoning maps for the UK. Nottingham, UK, British Geological Survey, Report, 62pp. (CR/07/125N).

Nicolson, H., Curtis, A., Baptie, B., 2014. Rayleigh Wave Tomography of the British Isles from Ambient Seismic Noise. Geophysical Journal International, 198, pp.637–655.

Patton, M., Shaw, R. & Petavratzi, E., 2025. Critical minerals and the circular economy in Northern Ireland, British Geological Survey Briefing Note

Pavey, A., Timms, R., Hampson, M., Jones, N., Shevelan, J., Coleman, C., 2025. 3D geological modelling to support an environmental safety case in the nuclear sector. Geological Society, London, Engineering Geology Special Publications, 30, 79-92

Pennington, C., Freeborough, K., Dashwood, C., Dijkstra, T., Lawrie, K., 2015. The National Landslide Database of Great Britain: Acquisition, communication and the role of social media. Geomorphology, 249, pp. 44-51.

Reuss, M., 2015. The use of borehole thermal energy storage (BTES) systems. In Woodhead Publishing Series in Energy, *Advances in Thermal Energy Storage Systems*, Woodhead Publishing, Luisa F. Cabeza (ed.), pp. 117-147, ISBN 9781782420880

Roy, C., Nowacki, A., Zhang, X., Curtis, A., Baptie, B., 2021. Accounting for natural uncertainty within monitoring systems for induced seismicity based on earthquake magnitudes. Frontiers in Earth Science, 9:634688.

Roy, C., Ryberg, T., Haberland C., Wellington K., Moynihan C., Rieger, P., Hitzman, M.W., 2026. On the usefulness of passive seismic imaging for mineral exploration - A case study in the Irish Midlands, Journal of Applied Geophysics, 244, 105991.

Scanlon, B.R., Fakhreddine, S., Rateb, A. et al., 2023. Global water resources and the role of groundwater in a resilient water future. Neture Reviews Earth & Envirnoment, 4, 87–101.

Smedley, P.L., Allen, G., Baptie, B.J., Fraser-Harris, A.P., Ward, R.S., Chambers, R.M., Gilfillan, S.M.V., Hall, J.A., Hughes, A.G., Manning, D.A.C., McDermott, C.I., Nagheli, S., Shaw, J.T., Werner, M.J., and Worrall, F., 2024. Equipping for risk: lessons learnt from the UK shale-gas experience on assessing environmental risks for the future geoenergy use of the deep subsurface. Science of The Total Environment, 921.

Stuart, M. E., 2011. Potential groundwater impact from exploitation of shale gas in the UK. British Geological Survey, Open Report OR/12/001. 33pp.

Strutz, D., and Curtis, A., 2025. The Roles of Low-Noise Stations, Arrays and Ocean-Bottom Seismometers in Monitoring UK Offshore Seismicity associated with Subsurface Storage of Carbon Dioxide. International Journal of Greenhouse Gas Control, 148, 104536

Strutz, D., Kiers, T., Curtis, A., 2025. Single and Multi-Objective Optimization of Distributed Acoustic Sensing Cable Layouts for Geophysical Applications. arXiv. doi: arXiv:2510.07531v1

TNO. 2025. Annual Report 2024: Natural Resources and Geothermal Energy in the Netherlands. Den Haag, Ministry of Climate and Green Growth, https://www.nlog.nl/media/3581.

Tromans, I.J., Aldama-Bustos, G., Douglas, J. *et al*. Probabilistic seismic hazard assessment for a new-build nuclear power plant site in the UK. Bulletin of Earthquake Engineering, 17, 1–36 (2019).

Tweed C., Ellis A., Whittleston R., 2015. Delivering safe geological disposal of nuclear waste in the UK. Proceedings of the Institution of Civil Engineers - Energy, 168, No. 4, pp. 206–217, doi: https://doi.org/10.1680/jener.14.00004

Verdon J.P., J-M. Kendall, A.L. Stork, R.A. Chadwick, D.J. White, R.C. Bissell, 2013. A comparison of geomechanical deformation induced by 'megatonne' scale CO2 storage at Sleipner, Weyburn and In Salah: Proceedings of the National Academy of Sciences 110(30), E2762-E2771.

Verdon, J.P., R. Schultz, B. Edwards, 2025. Tolerable magnitudes for induced seismicity from offshore carbon capture and sequestration projects around the United Kingdom: International Journal of Greenhouse Gas Control 142, 104335.

Watkins, T.J.M., Verdon, J.P., Rodríguez-Pradilla, G., 2023. The temporal evolution of induced seismicity sequences generated by low-pressure, long-term fluid injection, Journal of Seismology, 27, 243–259.

Williams, J. D. O., Williamson, J. P., Parkes, D., Evans, D. J., Kirk, K. L., Sunny, N., Hough, E., Vosper, H., Akhurst, M. C., 2022. Does the United Kingdom have sufficient geological storage capacity to support a hydrogen economy? Estimating the salt cavern storage potential of bedded halite formations. Journal of Energy Storage, 53, 105109.

Zhao, X., Lily Irvin, L., Galetti, E., Curtis, A., 2026. Direct-3D Variational Bayesian Surface Wave Inversion and Its Application to Ambient Noise Tomography beneath Great Britain. Geophysical Journal International, *submitted*; preprint available at: arXiv:2507.15390v2